\documentclass[numberedappendix]{emulateapj}

\begin{document}

\bibliographystyle{apj}

\shorttitle{Vega Debris Disk Structure}

\shortauthors{Hughes et al.}

\slugcomment{Accepted for Publication in ApJ: February 27, 2012}

\title{
Confirming the Primarily Smooth Structure of the Vega Debris Disk at Millimeter Wavelengths
}

\author{A. Meredith Hughes\altaffilmark{1,2},
David J. Wilner\altaffilmark{3},
Brian Mason \altaffilmark{4},
John M. Carpenter \altaffilmark{5},
Richard Plambeck \altaffilmark{1},
Hsin-Fang Chiang \altaffilmark{6,7},
Sean M. Andrews\altaffilmark{3},
Jonathan P. Williams \altaffilmark{8},
Antonio Hales \altaffilmark{9},
Kate Su \altaffilmark{10},
Eugene Chiang\altaffilmark{1},
Simon Dicker\altaffilmark{11}, 
Phil Korngut\altaffilmark{11}, 
Mark Devlin\altaffilmark{11}
}
\altaffiltext{1}{Department of Astronomy, University of California, Berkeley,
94720, USA; mhughes$@$astro.berkeley.edu}
\altaffiltext{2}{Miller Fellow}
\altaffiltext{3}{Harvard-Smithsonian Center for Astrophysics, 60 Garden Street, Cambridge, MA 02138, USA}
\altaffiltext{4}{National Radio Astronomy Observatory, 520 Edgemont Road, 
Charlottesville, VA 22903-2475, USA}
\altaffiltext{5}{California Institute of Technology, Department of Astronomy, MC 105-24, Pasadena, CA 91125, USA}
\altaffiltext{6}{Institute for Astronomy, University of Hawaii, 640 N. Aohoku 
Place, Hilo, HI 96720}
\altaffiltext{7}{Department of Astronomy, University of Illinois at Urbana-Champaign, 1002 West Green Street, Urbana, IL 61801, USA}
\altaffiltext{8}{Institute for Astronomy, University of Hawaii, 2680 Woodlawn
Dr., Honolulu, HI 96822, USA}
\altaffiltext{9}{Joint ALMA Observatory, Av. El Golf 40, Piso 18, Santiago, Chile}
\altaffiltext{10}{Steward Observatory, University of Arizona, 933 North Cherry Avenue, Tucson, AZ 85721, USA}
\altaffiltext{11}{Department of Physics and Astronomy, University of Pennsylvania, 209 S. 33rd St., Philadelphia, PA 19104, USA}

\begin{abstract}

Clumpy structure in the debris disk around Vega has been previously
reported at millimeter wavelengths and attributed to concentrations of
dust grains trapped in resonances with an unseen planet.  However, recent 
imaging at similar wavelengths with higher sensitivity has disputed the 
observed structure.  We present three new millimeter-wavelength observations 
that help to resolve the puzzling and contradictory observations.  We have 
observed the Vega system with the Submillimeter Array (SMA) at a wavelength 
of 880\,$\mu$m and angular resolution of 5''; with the Combined Array for 
Research in Millimeter-wave Astronomy (CARMA) at a wavelength of 1.3\,mm and 
angular resolution of 5''; and with the Green Bank Telescope (GBT) at a 
wavelength of 3.3\,mm and angular resolution of 10''.  Despite high sensitivity 
and short baselines, we do not detect the Vega debris disk in either of the 
interferometric data sets (SMA and CARMA), which should be sensitive at high 
significance to clumpy structure based on previously reported observations.  
We obtain a marginal (3$\sigma$) detection of disk emission in the GBT data; 
the spatial distribution of the emission is not well constrained.  We analyze 
the observations in the context of several different models, demonstrating 
that the observations are consistent with a smooth, broad, 
axisymmetric disk with inner radius 20-100\,AU and width $\gtrsim 50$\,AU.  
The interferometric data require that at least half of the 860\,$\mu$m 
emission detected by previous single-dish observations with the James Clerk 
Maxwell Telescope be distributed axisymmetrically, ruling out strong 
contributions from flux concentrations on spatial scales of $\lesssim$100\,AU.  
These observations support recent results from the Plateau de Bure 
Interferometer indicating that previous detections of clumpy structure in 
the Vega debris disk were spurious.

\end{abstract}
\keywords{circumstellar matter --- planetary systems --- planet-disk interactions --- stars: individual (Vega)}

\section{Introduction}

The presence of tenuous, second generation dust disks around main sequence
stars came as a surprise after the launch of the {\it Infrared Astronomical 
Satellite (IRAS)} in 1983 \citep{aum84}.  A variety of physical processes
is expected to remove orbiting dust grains on timescales shorter than the 
stellar age, primarily through collisional grinding and subsequent ejection 
of small grains by stellar radiation \citep[e.g.,][]{wya05,str06}.  
It is therefore thought that the circumstellar dust must be regenerated, 
presumably through grinding collisions of planetesimals.  Surveys for excess 
emission in the infrared have demonstrated that at least 15\% of nearby main 
sequence stars host debris disks \citep{hab01,bry06}.  As the sensitivity of
instruments improves, that fraction is steadily increasing.  Already the {\it 
Herschel} astronomical observatory has identified several new debris disk 
systems \citep[e.g.,][]{eir10,eir11,tho10}, with more likely to be announced 
soon as key programs wind down.  Spatially resolving the emission from the
extended, faint dust disks is challenging, however, and so far only a handful 
of systems out of hundreds of candidates have been spatially resolved.

Resolved observations of debris disks reveal a wide range of morphologies,
including broad disks, narrow rings, eccentricities, warps, and brightness 
asymmetries \citep[e.g.][]{hol98,sch99,wya99,hea00,kal05}.  Several of these 
morphological features have now been linked to the presence of 
planets in the disk; in fact, two of the three (so far) directly imaged 
extrasolar planetary systems were predicted based on the properties of their 
dust disks \citep[Fomalhaut and $\beta$ Pictoris;][]{kal08,lag10}.  That said, 
the clumpy structure predicted to result from large dust grains tracing 
planetary resonances has a more checkered history.  Theoretical predictions 
for the magnitude of dust concentration in orbital resonances vary widely, with
some groups predicting a pronounced contrast in the surface density of large
particles \citep[e.g.,][hereafter W03]{oze00,wya03} and others predicting very 
little structure even in the presence of massive planets 
\citep[e.g.,][]{kuc10}.  A complicating factor is that disk structure is 
predicted -- and, indeed, observed -- to vary strongly as a function of 
wavelength.  For example, several debris disks have now been observed to host 
smooth, extended haloes of small dust grains at radii far larger than the 
outer disk edges seen at longer wavelengths \citep[e.g.][]{su05,su09}.  

Millimeter-wavelength observations are preferred for tracing dynamical 
interactions with planets, since they trace the largest accessible 
dust grain populations that are the least sensitive to the smoothing effects
of stellar radiation \citep{wya06}.  Early observations were promising.  The 
Vega debris disk, in particular, exhibited a possible azimuthal asymmetry in 
the first 860\,$\mu$m James Clerk Maxwell Telescope (JCMT) map of the disk 
structure \citep{hol98}, which was supported by interferometric observations 
with the Owens Valley Radio Observatory \citep[OVRO;][]{koe01} and the 
Plateau de Bure Interferometer \citep[PdBI;][]{wil02}.  This result was 
followed by other claims of clumpy structure at millimeter 
wavelengths, including a suggestive observation of $\epsilon$ Eridani 
in which the clumps appeared to move at super-Keplerian velocities between 
epochs, as expected for a resonant orbital configuration \citep{gre05}.  
However, it was later pointed out that background galaxies may have contributed
to the apparent asymmetry and rotation rate in $\epsilon$ Eridani \citep{pou06},
and observations at a wavelength of 350\,$\mu$m did not confirm the presence 
of the 850\,$\mu$m clumps \citep{bac09}.  More recently, clumpy structure has 
been claimed (and planetary properties inferred) in millimeter-wavelength 
maps of the disks around the young solar analogue HD 107146 \citep{cor09} 
and the multiple-planet host star HR 8799 \citep{pat11}.  Unfortunately, 
more sensitive follow-up observations did not recover the claimed clumpy 
structure \citep{hug11}.  At the time of writing of this article, the 
observational evidence for clumpy millimeter-wavelength structure in debris 
disks is shaky at best.  However, it should be noted that {\it all} 
millimeter-wavelength observations of debris disks to date have been 
plagued by low signal-to-noise, due to the tenuous nature of 
the dust emission (which rarely amounts to more than an earth mass of 
material).  It is likely that sensitive future observations, e.g. with 
the Atacama Large Millimeter Array (ALMA), will uncover any resonant structure 
at lower contrast than the all-or-nothing levels that are currently 
observable.

The Vega debris disk provides an interesting case -- and is in many
ways the prototype -- of the study of clumpy structure in debris disks.  
The star itself is, of course, remarkably prominent both in the astronomical 
literature and in the night sky; it is an A0V star viewed nearly face-on to
its rotation axis \citep{gul94}, located at a distance of 7.7\,pc 
from the Sun \citep{per97,lee07}. Its age is estimated at 350\,Myr 
\citep{son00}, and it is host to the first extrasolar debris disk ever 
observed \citep{aum84}.  Attempts to infer the disk size from {\it IRAS} 
observations resulted in estimates of $\sim$20'' 
\citep[150\,AU;][]{har84,bli94}; subsequent {\it ISO}
data suggested that the disk may be even larger, up to 36'' (280\,AU) at 
a wavelength of 90\,$\mu$m \citep{hei98}.  The initial reconnaissance of disk 
morphology took place in the millimeter, with apparent clumpy structure on 
9'' (70\,AU) scales attributed to the presence of an orbiting 
Neptune-mass planet \citep[][W03]{hol98,koe01,wil02}.  Near-infrared 
interferometry with the Palomar Testbed Interferometer revealed a dust disk 
component much closer to the star, at a distance of only 4\,AU, recently 
confirmed with IOTA \citep{cia01,def11}.  After the flurry of work on the 
millimeter structure, it was with some surprise that \citet{su05} reported 
azimuthally symmetric dust emission observed with the {\it Spitzer} space 
telescope extending out to hundreds of AU at mid-IR wavelengths.  The smooth 
structure was confirmed by subsequent {\it Herschel} observations at 
wavelengths of 70--500\,$\mu$m; while the ring structure was resolved only 
at the shortest wavelengths, once again no azimuthal asymmetry was evident 
\citep{sib10}.  Meanwhile, Caltech Submillimeter Observatory (CSO) 
observations at wavelengths of 350 and 450\,$\mu$m resolved the ring structure 
and reported tentative evidence of clumpy structure in the disk \citep{mar06}.  
However, the position of the clumps in the CSO maps did not correspond with 
the clumps reported in the interferometric data; the authors suggested that 
perhaps the shorter-wavelength data were tracing 4:3 or 3:2 resonances rather 
than the 2:1 resonances detected at 860\,$\mu$m and 1.3\,mm wavelengths, and 
that two different grain populations may be responsible for the emission.  
However, the interferometric images were recently called into question by 
\citet{pie11}, who re-observed Vega with PdBI in a mosaicked observation at 
a factor of two higher sensitivity than before and did not recover any clumpy 
structure.

We present three new sets of observations of the Vega system at millimeter
wavelengths, with the intention of helping to resolve the puzzling and 
contradictory observations and models.  The data were obtained with the
Submillimeter Array (SMA)\footnote{The Submillimeter Array is a joint project between the Smithsonian Astrophysical Observatory and the Academia Sinica Institute of Astronomy and Astrophysics and is funded by the Smithsonian Institution and the Academia Sinica.}, 
the Combined Array for Research in Millimeter-wave Astronomy (CARMA)\footnote{Support for CARMA construction was derived from the Gordon and Betty Moore Foundation, the Kenneth T. and Eileen L. Norris Foundation, the James S. McDonnell Foundation, the Associates of the California Institute of Technology, the University of Chicago, the states of California, Illinois, and Maryland, and the National Science Foundation. Ongoing CARMA development and operations are supported by the National Science Foundation under a cooperative agreement, and by the CARMA partner universities. }, 
and the Green Bank Telescope (GBT)\footnote{The National Radio Astronomy Observatory is a facility of the National Science Foundation operated under cooperative agreement by Associated Universities, Inc.}, 
at wavelengths of 
880\,$\mu$m, 1.3\,mm, and 3.3\,mm, respectively.  We describe the observations 
in Section \ref{sec:obs}, present the results in Section \ref{sec:results}, 
and analyze the data in the context of several historically relevant classes 
of models in Section~\ref{sec:analysis}.  We discuss the results and summarize 
their significance in Section~\ref{sec:discussion}.

\begin{table*}
\caption{Basic Observing Parameters}
\begin{tabular}{lccc}
\hline
Date & 225\,GHz opacity & Track length (hr) & RMS Sensitivity (mJy\,beam$^{-1}$) \\
\hline
\multicolumn{4}{c}{341\,GHz SMA - 4\,GHz bandwidth} \\
\hline
2007 Apr 20 & 0.05 & 8 & 2.6 \\
2007 Apr 21 & 0.05-0.06 & 7 & 2.6 \\
2007 Apr 22 & 0.04 & 8 & 2.2 \\
2007 Apr 23 & 0.03 & 8 & 2.0\\
\hline
\multicolumn{4}{c}{227\,GHz CARMA - 8\,GHz bandwidth} \\
\hline
2011 Jan 20 & 0.15 & 3 & 0.54 \\
2011 Jan 23 & 0.10 & 5 & 0.53 \\
2011 Jan 30 & 0.25 & 4 & 0.87 \\
2011 Feb 2 & 0.07 & 2 & 0.87 \\
2011 Feb 7 & 0.15 & 3 & 0.71 \\
2011 Feb 8 & 0.14 & 6 & 0.61 \\
2011 Feb 9 & 0.13 & 6 & 0.51 \\
\hline
\end{tabular}
\tablenotetext{a}{Angular resolution using natural weighting of visibilities.}
\label{tab:obs}
\end{table*}

\section{Observations}
\label{sec:obs}

\begin{figure}[t]
\epsscale{1.0}
\plotone{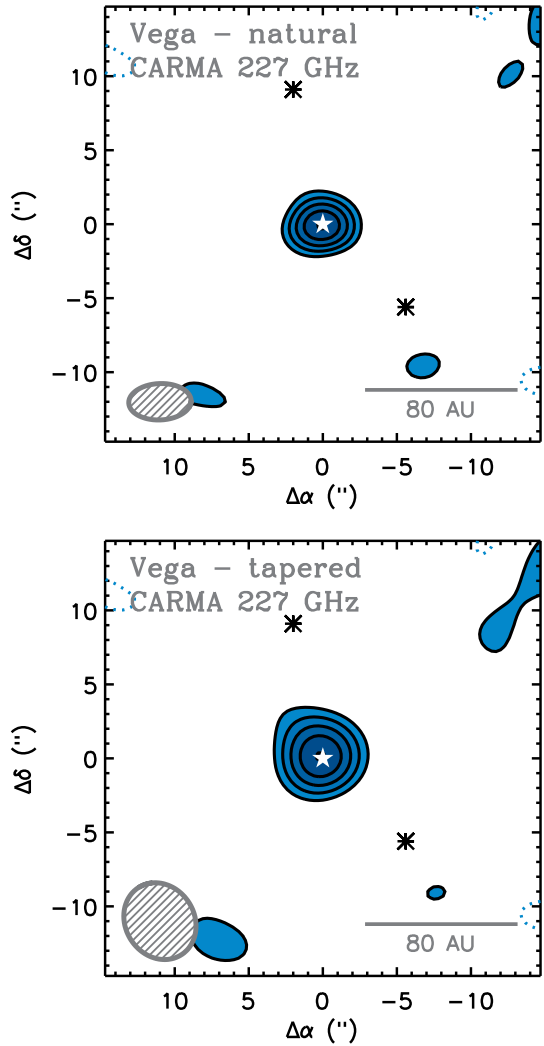}
\figcaption{CARMA maps of the 227\,GHz continuum emission from the Vega system.
The star symbol marks the position of Vega; thermal emission from the stellar
photosphere is detected at the 6$\sigma$ level.  The top panel shows an image
with naturally weighted visibilities, resulting in a beam size of 
4\farcs2$\times$2\farcs5 with position angle -88$^\circ$.  The bottom panel has
been tapered in the visibility domain to match the spatial resolution of 
Figure 1c of \citet{wil02}, resulting in a beam size of 
5\farcs4$\times$4\farcs7 at a position angle of 32$^\circ$.  The contours are 
[2,3,4,5]$\times$0.28\,mJy\,beam$^{-1}$ (the rms noise).  The asterisk symbols (along 
with the star symbol) mark the pointing centers of the 3-point mosaic, and 
are chosen to coincide with the position of the emission peaks reported in 
\citet{wil02}.  No thermal emission from the dusty disk is detected, despite 
shorter baselines and lower rms noise than the PdBI data set.
\label{fig:carma}}
\end{figure}

Data were collected in a series of 8 observing sessions on the GBT
in spring 2009 and spring 2010.  The GBT's 90 GHz bolometer array,
MUSTANG \citep{dic09} was used for these observations, giving a
$9"$ FWHM beam on the sky.  A total of $9.4$ hours of useful integration
time were acquired on the map of Vega.  A 1.5$\times$1.5 arcmin map around 
Vega was obtained with alternating Lissajous patterns, some centered at the 
position of the star and four others offset by $\pm$15'' in both north-south
and east-west directions.  Primary flux calibration was with reference to 
the asteroid Ceres-1, using a light curve and brightness temperature 
provided by T. Mueller (private comm).  Every 30-45 minutes the GBT beam, 
focus, and pointing was checked by a short (1 minute) observation of a 
compact pointing calibration source within 15 to 20 degrees of Vega. 
Pointing corrections were applied offline in the data analysis. If the 
beam or focus showed evidence of drift from their optimal values, they 
were corrected by collecting a 15-minute Out-of-focus (OOF) holography 
map, analyzing the data on the fly, and applying the corrections to the 
telescope \citep{nik07}. Noise levels were established by means of a suite 
of Monte-Carlo bootstrap simulations in which the bolometer data were shifted 
by random amounts in time with respect to the antenna position data. Images 
were made by a series of IDL routines developed to analyze MUSTANG data. 
These routines remove a common mode from all detectors, which is highly 
effective at eliminating large-scale ($>$40'') signals due to fluctuations 
in the atmosphere emission and cryogenic instabilities. A low-order polynomial 
is fitted to each detector to remove residual drifts; the individual detector 
weights are calculated from the RMS of the residual timestream; and the 
resulting data gridded onto the sky.  More details about the data reduction 
and calibration procedures can be found in \citet{mas10} and \citet{dic09}.

The SMA data were collected during four eight-hour tracks on 20-23 April 
2007 in the subcompact configuration, which provided baseline lengths of 10 
to 45\,m between the eight six-meter antennas (see Table~\ref{tab:obs} for 
basic information about individual tracks). 
The weather was excellent, particularly on the night of April 23.  The 
225\,GHz opacity was stable between 0.04 and 0.06, with the exception of a 
few spikes in humidity on the night of April 21.  The correlator was 
configured for maximum continuum sensitivity, with uniform spacing of 32 
channels in each 104\,MHz correlator chunk across the 2\,GHz bandwidth in 
each sideband.  The LO frequency was set to 340.8\,GHz, placing the CO(3-2) 
line in the upper sideband.  Flux calibration was carried out using Titan 
and MWC 349.  Because of the range in right ascension covered by the target 
and calibrators, atmospheric and instrumental gain calibration was carried 
out using both MWC 349 and 3c345, with the former dominating the early part 
of the track and the latter the late part of the track.  The derived flux of 
MWC 349 was 2.45\,Jy; 3c345 varied between 2.29 and 2.51\,Jy over the four 
nights.  The quasar J1801+440 was included in the observing loop to test 
the quality of the phase transfer to Vega, although the photosphere of Vega 
was also detectable on all four nights and served as a secondary check on 
the calibration. Errors in the nominal baseline solutions were evident in 
the quasar phases, which did not track each other well.  However, we were able 
to derive corrected baseline solutions using only the calibrators present 
in the data set.  The updated baseline solutions resulted in a 10\% 
improvement in signal-to-noise on the detection of Vega's photosphere.
Routine calibration tasks were carried out using the MIR\footnote{See
http://cfa-www.harvard.edu/$\sim$cqi/mircook.html} software package, while
imaging and deconvolution were accomplished using the MIRIAD software package.
A naturally weighted image of the four data sets yields an rms noise of 
0.94\,mJy\,beam$^{-1}$ in a 5\farcs3$\times$4\farcs7 beam at a position angle
of -80$^\circ$.

The CARMA data were collected in a series of seven short (3-6 hour) tracks 
between 20 January and 9 February 2011 (see Table~\ref{tab:obs}).  The 
array was in its most compact ``E'' configuration, with projected baseline 
lengths between 5 and 65\,m.  The shortest baselines were provided by the 
nine 6\,m diameter antennas, while the longest baselines were preferentially 
between the six 10\,m diameter antennas.  The correlator was configured for 
maximum bandwidth and therefore continuum sensitivity, with eight 500\,MHz 
bands in each sideband for a total bandwidth of 8\,GHz.  The LO frequency 
was 227.5\,GHz and the IFs of the bands were distributed between 2.3 and 
7.3\,GHz in such a way as to avoid atmospheric ozone lines.  The observing 
pattern was a three-point mosaic centered on the star position and the 
positions of the two clumps reported in \citet{wil02}.  Observations of Vega 
alternated with the quasar J1848+323, which was used to calibrate the 
atmospheric and instrumental phase variations.  No test quasar was included 
in the observing loop since Vega's photosphere was bright enough to be 
detected in most of the tracks (all but the two with the poorest observing 
conditions) and provided an excellent test of the quality of the phase 
transfer.  The quasar 3c454.3 was used as the passband calibrator, and Neptune 
as the flux calibrator.  The derived flux of J1848+323 varied between 0.45 
and 0.63\,Jy, yielding a photospheric flux for Vega of approximately 
0.7\,mJy\,beam$^{-1}$.  Because Vega was a daytime source in January and 
the pointing on the 6\,m antennas is poorer during the day, optical pointing 
was carried out every two hours, using the photosphere of Vega as the 
pointing calibrator \citep{cor10}.  Calibration, imaging, and deconvolution 
were accomplished using the MIRIAD software package.  The rms noise in the
naturally weighted image combining all the data was 0.28\,mJy\,beam$^{-1}$ in
a 4\farcs2$\times$2\farcs5 beam at a position angle of -88$^\circ$.

\section{Results}
\label{sec:results}

We detect the stellar photosphere in each of the three data sets, but only in
the GBT data is there a marginal (3$\sigma$) detection of thermal emission 
from the dust disk.  The disk emission is not detected in either of the two
interferometric data sets despite lower rms noise and shorter baselines than
previous observations that apparently showed clumpy dust emission.

Fig.~\ref{fig:carma} shows the results of the CARMA observations.  The stellar
photsphere is independently detected in most of the seven short tracks, and 
the photospheric detection for the combined data set is at the 6$\sigma$ level.
The top panel of Fig.~\ref{fig:carma} shows the combined data set imaged with
natural weighting, while the bottom panel shows the same visibilities imaged 
with a Gaussian taper to recover the lower spatial resolution with which the
clumps were previously claimed to be seen.  The asterisk symbols in the two 
panels mark the pointing centers of the 3-point mosaic (with the third point 
located at the star position), chosen to coincide with the location of the 
putative clumps in the \citet{wil02} data.  No emission is detected at those 
locations in the CARMA data set despite a factor of two lower rms noise.  
The observations also have lower rms noise and richer ($u$,$v$) coverage 
than the OVRO data obtained by \citet{koe01} using the 10.4\,m antennas that 
have since been incorporated into CARMA.  

\begin{figure}[t]
\epsscale{1.0}
\plotone{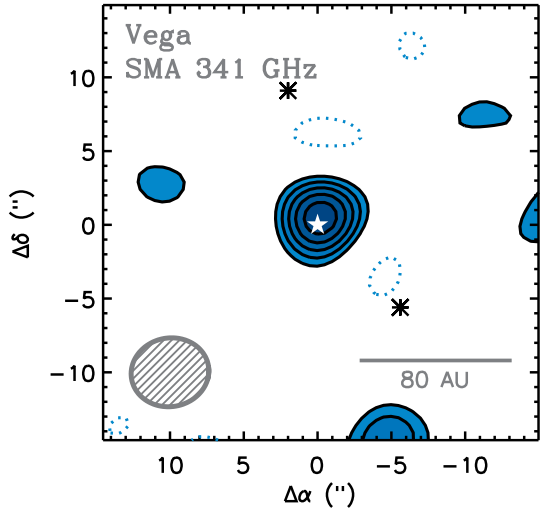}
\figcaption{SMA map of the 341\,GHz continuum emission from the Vega system.
Thermal emission from the stellar photosphere is detected at the 7$\sigma$ 
level.  The visibilities have been imaged using natural weighting, resulting 
in a beam size of 4\farcs7$\times$5\farcs3 with position angle -80$^\circ$.  
The contours are [2,3,4,...]$\times$0.94\,mJy\,beam$^{-1}$ (the rms noise).  Symbols are
as in Fig.~\ref{fig:carma}.  No thermal emission from the dusty disk is
detected, consistent with the CARMA (Fig.~\ref{fig:carma}) and PdBI results 
\citep{pie11}.
\label{fig:sma}}
\end{figure}

The SMA observations in Fig.~\ref{fig:sma} tell a similar story.  
The photosphere is pointlike, implying successful calibration of 
the instrumental and atmospheric gains, and is strongly detected at the 
7$\sigma$ level.  The higher frequency of the SMA observations requires 
flux scaling for comparison to the 1.3\,mm data.  Assuming a 
millimeter-wavelength spectral index of 2.8 for the disk, consistent with
the long-wavelength {\it Herschel} fluxes reported in \citet{sib10}, the 
expected integrated 880\,$\mu$m flux of the emission peaks reported in 
\citep{wil02} is 20 and 12\,mJy for the northeast and southwest peak, 
respectively, with peak fluxes at the resolution of the data of approximately 
7.4\,mJy.  Given the 0.9\,mJy rms noise in the data, we would expect to detect 
emission from the clumpy dust disk at the 8$\sigma$ level (similar to the
detection level of the stellar photosphere).  However, no such emission was
observed.  This is comparable to the limits set by the CARMA data 
(Fig.~\ref{fig:carma}) and the recent updated observations from the Plateau 
de Bure Interferometer \citep{pie11}.  

Taken together, these
three interferometric data sets, observed with three different instruments
with broad ($u$,$v$) coverage and uniformly low rms noise, place strong 
constraints on the amount of concentrated, clumpy emission that might be
present in the Vega debris disk.  They also demonstrate that previous
detections of clumpy emission from the Vega system using interferometers 
must have been spurious.  The noise may have been difficult to characterize 
near the edges of the primary beam, a problem mitigated in the recent data 
sets by the use of mosaicked imaging (CARMA, PdBI) and smaller antennas (SMA).  

\begin{figure}[t]
\epsscale{1.0}
\plotone{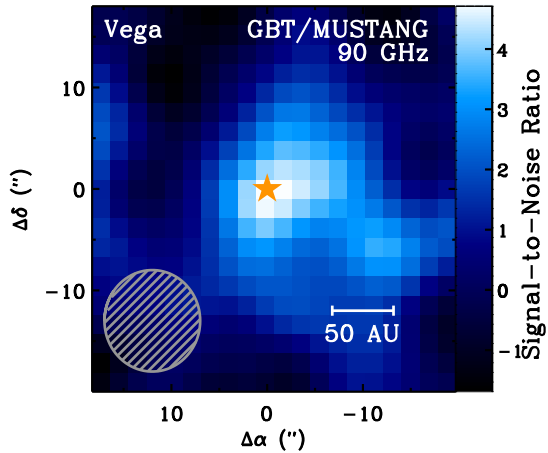}
\figcaption{90\,GHz GBT/MUSTANG map of continuum emission from the Vega 
system.  The GBT map has been smoothed to 10'' resolution (with 2'' pixels), 
indicated by the beam in the lower left corner.  The rms noise varies with 
position due to the scan pattern, with the lowest rms closest to the pointing 
center, so the data are presented as a signal-to-noise map to emphasize the 
significance of the emission detection as a function of position on the sky.  
The emission is centered on the star position (peak signal-to-noise ratio of 
4.7), and the integrated flux is 690$\pm$130\,$\mu$Jy within a 16'' radius 
from the star, indicating a marginal (3\,$\sigma$) detection of emission 
above the 310\,$\mu$Jy photosphere.
\label{fig:gbt}}
\end{figure}

\begin{figure}[t]
\epsscale{1.0}
\plotone{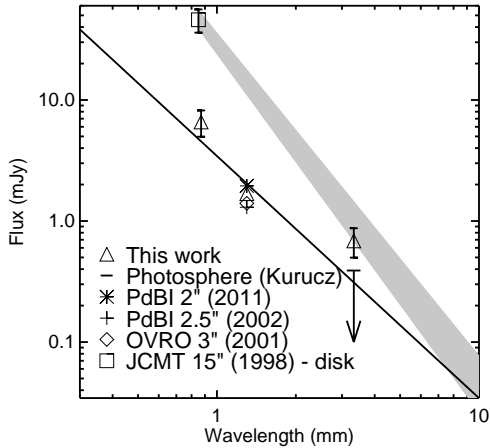}
\figcaption{Millimeter-wavelength SED of Vega.  Flux values 
from this work are plotted with triangle symbols.  Error bars mark 1-$\sigma$ 
intervals, including both the noise in the image and a 20\% systematic 
flux uncertainty.  Error bars are not included on the 1.3\,mm data to avoid 
excessive cluttering around the four closely spaced points.  The telescope 
name, angular resolution, and year of observation is indicated in the legend.
Data are drawn from \citet{hol98}, \citet{koe01}, \citet{wil02}, and 
\citet{pie11}.  The disk is resolved out in the interferometric data (all 
points except the 860\,$\mu$m JCMT flux from Holland et al. 1998 and the 
3.3\,mm GBT flux), which are sensitive only to the photosphere.  A 3.3\,mm 
upper limit (3$\sigma$) on the photosphere from \citet{wil02} is also 
included.  The gray area represents an extrapolation of the 46$\pm$5\,mJy 
integrated 860\,$\mu$m flux measured with the JCMT \citep{hol98} to longer 
wavelengths, assuming a spectral index of 2.8$\pm$0.1 estimated from fitting 
the long-wavelength {\it Herschel} data \citep{sib10}.  An extrapolation of 
the 2005 Kurucz model photosphere (solid line) is also plotted.
\label{fig:sed}}
\end{figure}

Given the smoothness of the disk emission indicated by the new interferometric 
observations, single-dish data are more desirable for characterizing the 
amount and location of emission from the dusty disk.  The 90\,GHz GBT map is
presented in Fig.~\ref{fig:gbt}.  The star position accounts for proper motion 
and is offset from the map pointing center by a few arcseconds.  The location 
of the emission peak coincides with the star position, although it is only 
determined to within 3'' or so due to pointing uncertainties.  The map has 
been smoothed to 10'' resolution (with 2'' pixels).  Because of the scanning 
pattern used to minimize atmospheric effects, the rms noise varies with 
position; it is lowest near the map center and highest at large distances 
from the map center.  Fig.~\ref{fig:gbt} shows a $\pm$20'' box around the 
star position.  The emission peaks near the star position with a maximum 
signal-to-noise ratio of 4.7.  The integrated flux within a 16'' radius of 
the star is 690$\pm$130\,$\mu$Jy.  To determine the expected photospheric 
contribution at this wavelenghth, we use the most recent Kurucz model 
photosphere of Vega (R. Kurucz, priv. comm.), which is scaled to match the 
flux of 7.17\,Jy at 23.68\,$\mu$m derived by \citet{rie08}.  We extrapolate 
to the millimeter regime assuming that the flux scales as $\nu^2$ in the 
Rayleigh-Jeans tail of the distribution.  This yields photospheric fluxes 
of 4.4\,mJy, 2.1\,mJy, and 310\,$\mu$Jy at wavelengths of 880\,$\mu$m, 
1.3\,mm, and 3.3\,mm, respectively.  The GBT data therefore imply a marginal 
(3$\sigma$) detection of 375\,$\mu$Jy of emission above the photosphere, 
presumably from the dusty disk around Vega.  Assuming a characteristic 
dust temperature of 70\,K \citep{su05,su06} and a standard 90\,GHz dust 
opacity of 0.9\,cm$^2$\,g$_\mathrm{dust}^{-1}$ \citep{bec90}, this 
corresponds to a mass of 2.3$\times$10$^{-3}$\,M$_\earth$, in good agreement 
with more detailed SED-based models \citep{su05}.  Figure~\ref{fig:sed} 
plots the measured photospheric fluxes at 880$\mu$m and 1.3\,mm along with 
the Kurucz model photosphere extrapolation to the millimeter spectral regime.  
A 3.3\,mm upper limit (3$\sigma$) on the photospheric flux from \citet{wil02} 
is also plotted.  The gray shaded region of the plot represents an 
extrapolation of the total 860\,$\mu$m flux from the disk reported by 
\citet{hol98}, assuming a spectral index of 2.8$\pm$0.1 estimated from 
fitting the long-wavelength {\it Herschel} data \citep{sib10}.  The expected 
stellar contribution at 860\,$\mu$m is 4.4\,mJy, while extrapolation of 
the {\it Herschel} data predict a larger 860\,$\mu$m flux by about the same 
amount (which is within the measurement uncertainty).  We therefore assume 
that the disk flux at 860\,$\mu$m is 46\,mJy.  The GBT 3.3\,mm flux is only 
marginally consistent with both the Kurucz model photosphere extrapolation 
and the PdBI upper limit on the photospheric flux at 3.3\,mm \citep{wil02}, 
supporting the conclusion that the disk contributes to the measured 3.3\,mm 
flux.  A substantially flatter millimeter spectral index would weaken the 
inferred detection of 3\,mm emission from the Vega debris disk.

Given the low spatial resolution and signal-to-noise in the GBT data, it is 
difficult to draw strong conclusions about the spatial distribution of the 
emisson, although the implied integrated 90\,GHz emission from the dust disk 
within 16'' (125\,AU) is approximately 375\,$\mu$Jy.  This is consistent 
with previous observations from the JCMT by \citet{hol98}, who 
report a peak flux of 17.3$\pm$3.0\,mJy in a 14'' beam at a wavelength of 
860\,$\mu$m.  Scaling the 860\,$\mu$m flux to the 3.3\,mm wavelength of the 
GBT observations predicts 380\,$\mu$Jy of dust emission within 14'' of the 
star, using the spectral index of 2.8$\pm$0.1 derived from the long-wavelength 
{\it Herschel} data \citep{sib10}.  The integrated emission in the JCMT map 
was larger, 46$\pm$5\,mJy, but the increasing rms noise with distance from 
the GBT map center makes it difficult to detect very extended emission, so 
we limit ourselves to considering emission only within $\sim$125\,AU from 
the star.  Also, given the position uncertainty in the GBT map, it is 
difficult to determine whether the emission peaks at the star position.  
The 3'' uncertainty makes the GBT map at best marginally consistent with 
the 9'' offset of the peak emission in the JCMT map; however, the photosphere 
represents a much larger contribution to the total flux at 3.3\,mm than at 
860\,$\mu$m, which is likely to shift the location of the emission peak 
closer to the center as the disk emission is convolved with the photsphere.  
To summarize, while the GBT data are consistent with the JCMT map, 
on their own they can neither confirm nor disprove the presence of 
non-axisymmetric emission in the Vega debris disk.

\section{Analysis}
\label{sec:analysis}

The data presented in this paper, along with recent PdBI observations from
\citet{pie11}, represent a marked departure from previous observations 
that indicated clumpy emission from the Vega debris disk.  We therefore seek to 
characterize the parameter space consistent with the new observations.
We consider three categories of models: (1) the dynamical model of a resonant 
planetesimal population from W03, similar to that presented in 
\citet{wil02} and designed to reproduce the JCMT map; (2) a smooth, 
axisymmetric disk model, motivated by ring-like structure observed at shorter 
wavelengths; 
and (3) a ``toy'' model of 
clumpy debris disk emission that can be distilled into a two-parameter space 
to constrain exactly how clumpy (or axisymmetric) the disk emission may be 
while maintaining consistency with the data\footnote{We also tried an eccentric
ring model motivated by the double-peaked structure in the JCMT and GBT maps,
but it required emission too concentrated to be consistent with the SMA and 
CARMA data}.  We compare the data with these models in sequence in the 
following sections.  While the GBT data are not sensitive enough to strongly 
constrain the degree of axisymmetry of the disk emission at large distances 
from the star, they are useful for constraining the radial size of the flux 
distribution.  We therefore include these data in the subsequent analysis.

\subsection{Resonant Clump Model}
\label{sec:wyatt}

Since the W03 model was generated to reproduce the clumpy emission 
reported in the older data sets \citep{hol98,koe01,wil02}, it is likely
that it will be inconsistent with the more sensitive upper limits on 
clumpy disk emission from the PdBI \citep{pie11}, SMA, and CARMA (this work).  
Nevertheless, for completeness we present a comparison of the new SMA, CARMA
and GBT data with the W03 model, scaled to the appropriate
wavelengths.  

\begin{figure*}[t]
\epsscale{1.0}
\plotone{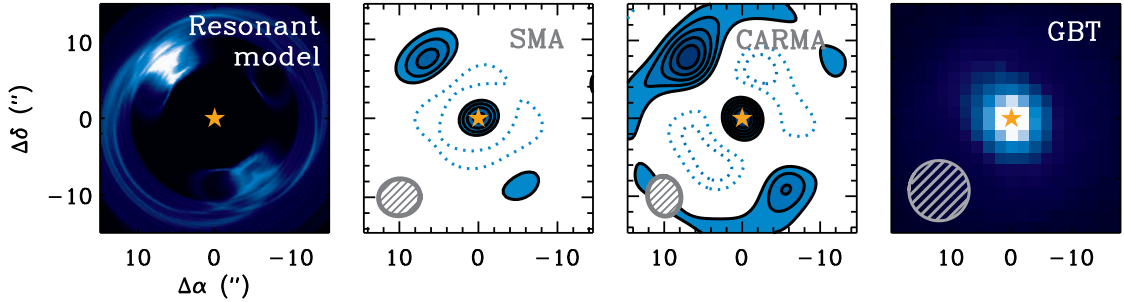}
\figcaption{Simulated observations of the W03 resonant model (left) 
with each of the different observing setups presented in this paper.  If the 
model were correct, we would expect to detect dust emission at the 4$\sigma$ 
level with the SMA at 341\,GHz (center left) and at the 6$\sigma$ level with 
CARMA at 227\,GHz (center right).  The 90\,GHz GBT observations are consistent 
with the W03 model primarily because they do not have the spatial 
resolution or sensitivity necessary to discriminate between this and other 
models.  The observations are simulated without noise and with identical 
imaging parameters and flux cutoffs to the images in 
Figs.~\ref{fig:carma} and \ref{fig:sma} (the negative residuals in the 
interferometric maps are artifacts of the CLEAN algorithm applied to 
the data), but the W03 model is clearly inconsistent with the 
observations, particularly the CARMA data.  Symbols are as in 
Figs.~\ref{fig:carma}-\ref{fig:gbt}, including SMA contour levels of 
[2,3,4,...]$\times$0.94\,mJy\,beam$^{-1}$ and CARMA contour levels of 
[2,3,4,...]$\times$0.28\,mJy\,beam$^{-1}$.  
\label{fig:wyatt}}
\end{figure*}

Figure~\ref{fig:wyatt} shows a comparison of the W03 model and (noiseless) 
simulations of each of the three data sets presented in this paper.  To 
simulate the interferometric observations, we sample the model at the same 
spatial frequencies as the data, then invert the visibilities and deconvolve 
the image with the same parameters as were used for the data, including the 
flux cutoff in the CLEAN procedure.  The negative contours in the 
interferometric images are artifacts of the CLEAN algorithm applied to 
the data.  The W03 model of the underlying planetesimal distribution is 
scaled to an appropriate flux at each wavelength.  At 341\,GHz, the 
photosphere is assumed to be 4.4\,mJy and the total flux in the disk model 
plus photosphere is scaled to match the integrated flux of 46\,mJy reported 
for the JCMT detection \citep{hol98}.  At 227\,GHz, the 850\,$\mu$m model is 
scaled in flux assuming a spectral index of 2 for the star and 2.8 for the 
disk emission.  The 90\,GHz model prediction is generated by assuming a 
310\,$\mu$Jy photosphere and scaling the flux within 16'' of the star position 
to the integrated flux in the GBT map.  The resulting simulated map 
demonstrates that the GBT data are not sensitive enough to rule out a 
resonant emission model.  While the SMA data are similarly consistent with 
the W03 simulations, the CARMA data are clearly inconsistent with the model 
prediction.

\subsection{Axisymmetric Model}
\label{sec:smooth}

Since a clumpy model is inconsistent with the observations, it is 
desirable to characterize the parameter space allowed for a smooth, 
axisymmetric model of the disk emission.  A smooth model is 
consistent with shorter-wavelength observations of the Vega debris disk 
\citep[e.g.,][]{su05} and with submillimeter 
observations from \citet{mar06}, and is perhaps the most natural explanation 
for the lack of detectable millimeter-wave emission in the interferometric 
data sets.  We therefore investigate the parameter space allowed by the 
combination of SMA, CARMA, and GBT data presented in this paper.  

We assume a smooth, axisymmetric disk with inner radius $R_\mathrm{in}$ and
width $\Delta R$ (so that the disk outer radius $R_\mathrm{out} = R_\mathrm{in}
+ \Delta R$).  We assume that the disk is viewed face-on ($i$=0).  The 
inclination of the stellar rotation axis to our line of sight is closer to 
5$^\circ$ \citep{auf06}, but the true disk inclination is not well determined.
Since many previous papers modeling the Vega debris disk have assumed a face-on
orientation for simplicity, we follow suit; a difference of a few degrees does 
not alter our conclusions.  We assume constant surface density between 
$R_\mathrm{in}$ and $R_\mathrm{out}$, with temperature decreasing as $T(
R) \propto R^{-0.5}$ and opacity constant with radius.  Effectively, this 
amounts to an assumption that flux also decreases with distance from the star 
as $R^{-0.5}$.  We make no assumptions about dust grain composition, emission 
or absorption efficiency, or opacity.  The temperature and mass of the dust 
are thoroughly degenerate, and we concern ourselves only with the distribution 
of {\it flux} in the disk, i.e., the product of surface density, temperature, 
and opacity.  For standard assumptions about the dust grain composition and 
opacity, the temperatures assumed in our model are consistent with those in 
the literature \citep[e.g.,][]{su05}.  We further adopt the observational 
constraints that the total flux of star and disk at 860\,$\mu$m is 45.7\,mJy 
\citep{hol98}; the photospheric flux at 341\,GHz is 5.7\,mJy; the flux at 
227\,GHz scales from 860\,$\mu$m with a spectral index of 2.8; and the total 
disk flux within 16'' from the star at 90\,GHz is 375\,$\mu$Jy.  Under these 
constraints, we vary $R_\mathrm{in}$ and $\Delta R$ to generate a grid of 
models to be compared with the data.  For the interferometric data sets, we 
generate a model of the disk only (without the photosphere), and sample the 
visibilities with the same spatial frequencies as the observations using the 
MIRIAD task \texttt{uvmodel}.  We then compare the synthetic visibilities with
the observed visibilities, real and imaginary, and calculate a $\chi^2$ value.  
For the GBT 
data, we create a model image of the photosphere and disk, convolve it with 
an appropriately sized beam, and calculate the reduced $\chi^2$ value in 
a 16'' square region around the star position, assuming the 75\,$\mu$Jy rms 
noise in the map center derived from simulated observations in 
Section~\ref{sec:obs}. 

\begin{figure}[t]
\epsscale{1.0}
\plotone{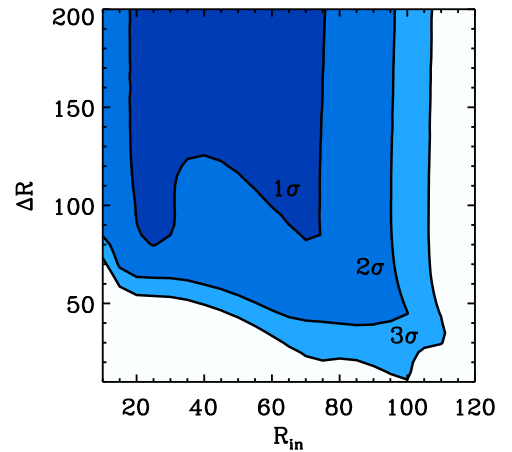}
\figcaption{Confidence interval contour plot showing the regions of parameter 
space for a smooth disk model allowed by the combination of SMA (5'' 
resolution), CARMA (4''), and GBT (10'') data.  The abscissa corresponds to 
the inner radius of an axisymmetric disk while the ordinate corresponds to 
the width of the disk (so that the disk outer radius 
$R_\mathrm{out}=R_\mathrm{in}+\Delta R$).  Small inner radii and small disk 
widths are ruled out by the interferometric data, while large inner radii 
are ruled out by the GBT data. 
\label{fig:smooth}}
\end{figure}

Fig.~\ref{fig:smooth} presents the results of the simulations, and marks off
the parameter space allowed by the SMA, CARMA, and GBT observations, given
the aforementioned assumptions.  Small values of $R_\mathrm{in}$ and $\Delta
R$ are ruled out by the interferometric data, since emission concentrated too
close to the star position or in too narrow a ring around the star should have
been detected in the CARMA or SMA data sets.  Models with larger 
$R_\mathrm{in}$ and $\Delta R$ exhibit smoother emission that would be 
undetectable by the interferometric observations; however, based on the GBT
detection of 375\,$\mu$m within 16'' (125\,AU) of the star, $R_\mathrm{in}$ 
must be smaller than this aperture.  The flux distribution in the GBT data 
places constraints on smaller radii as well.  In the context of a smooth, 
axisymmetric disk model, an inner radius of $\sim$20-100\,AU is preferred, 
with a width $\gtrsim$50\,AU.  Inner radii $<$15\,AU and $>$110\,AU are 
ruled out by the CARMA and GBT data, respectively, while widths $<$30\,AU 
are ruled out by the interferometric data.

\subsection{Clumpy Toy Model}

Although the data presented in \citet{pie11} and this work are consistent with
an azimuthally symmetric distribution of emission, it is still true that 
some millimeter data sets \citep{hol98,mar06} exhibit marginally significant 
evidence for a clumpy distribution of emission.  In an attempt to quantify 
exactly how clumpy the Vega debris disk may be given the available data, we 
consider a toy model consisting of a smooth component underlying two 
Gaussian ``clumps'' of variable FWHM.  We make several simplifying 
assumptions: (1) the location and relative brightness of the two clumps are 
fixed at 9'' northeast and southwest of the star position, respectively, with 
the southwest peak 0.3 times that of the northwest peak, consistent with a 
literal interpretation of the JCMT data; (2) the inner and outer radius of 
the smooth component are fixed at 60 and 200\,AU, respectively, to maintain 
consistency with the results of the axisymmetric modeling from 
Section~\ref{sec:smooth}; and (3) the surface density of the smooth component 
is held constant, while the temperature is assumed to decrease with distance 
from the star $R$ as $R^{-1/2}$.  As in Section~\ref{sec:smooth}, we also 
normalize the 880\,$\mu$m flux to the measured JCMT value from \citet{hol98}, 
assume a spectral index of 2.8 to extrapolate to 227\,GHz, and normalize the 
90\,GHz disk flux to the measured 375\,$\mu$Jy within 125\,AU from the star.  
Given these assumptions, the only variables are the fraction 
$f_\mathrm{clump}$ of total flux in the clumpy component (with the rest 
distributed across the smooth component), and the FWHM of the Gaussian 
clumps.  While we do fix the position of the clumps in the disk to match the 
JCMT observation, the result will be more generally applicable since the 
interferometric data would be capable of detecting comparably clumpy structure 
at virtually any position in the map.

\begin{figure}[t]
\epsscale{1.0}
\plotone{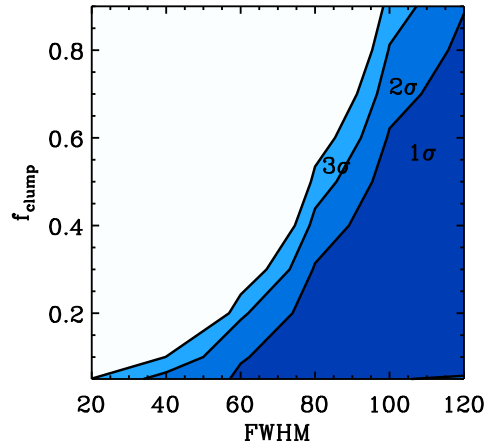}
\figcaption{Confidence interval contour plot showing the regions of parameter 
space for a clumpy disk model allowed by the interferometric (SMA and CARMA) 
data.  The abscissa corresponds to the FWHM of the Gaussian clumps in AU, while
the ordinate corresponds to the fraction of flux in the clumpy rather than
smooth component.  
\label{fig:clumpy}}
\end{figure}

We first examine only the constraints from the interferometric data, since 
these data are the most sensitive to the presence or absence of flux
concentrations.  Fig.~\ref{fig:clumpy} shows the parameter
space permitted by the SMA and CARMA observations: as expected, a smooth
brightness distribution is favored over a clumpy distribution, although very 
diffuse and extended clumps (with FWHM $\gtrsim 100$\,AU) are permitted.  It
should be noted that the largest size scale we consider is $\sim$120\,AU, 
since this is roughly the largest size scale to which the data are sensitive,
given the $\sim$1\,mm wavelength of observation and the $\sim$10\,m baselines
in the CARMA and SMA data.  However, even when diffuse clumps are included, 
they should not account for more than about half of the total flux in the image 
to maintain consistency with the interferometric data.  The constraints from 
the GBT data are somewhat orthogonal: due to the flux concentration in the 
southwest corner of the GBT map, the data favor models with a clumpy 
underlying flux distribution. The overlapping preferred parameter space for 
all three data sets occurs near the upper right corner of the plot, 
with large, diffuse clumps accounting for approximately half of the flux in 
the data, and the rest distributed in an extended smooth component.  However, 
considering the low signal-to-noise ratio for the extended (non-photospheric) 
flux in the GBT map, we note that the constraints from the interferometric 
map are the strongest, and require the smooth component of the flux 
distribution to dominate.  The CARMA and SMA therefore demonstrate that most 
($\gtrsim$50\%) of the flux is distributed in a smooth, axisymmetric ring 
around the star, although a small degree of concentration on very large scales 
is permitted.  This result does not depend strongly on the assumed values of 
R$_\mathrm{in}$ and $\Delta$R.

\section{Discussion}
\label{sec:discussion}

Modeling of the three new data sets favors a smooth, axisymmetric flux 
distribution for the Vega debris disk, with an inner radius of 20-100\,AU 
and a broad radial width of at least 50\,AU.  Diffuse emission on large scales 
may account for up to about half of the flux, but the majority must be 
smoothly distributed.  We have considered only a small subset of possible 
flux configurations for the disk, but the results are more generally 
applicable.  In particular, the clumpy toy model demonstrates that any flux 
concentration on spatial scales of less than $\sim$100\,AU cannot dominate the 
millimeter-wavelength flux distribution.  While we chose a two-clump model 
to approximate the morphology of the JCMT data, the result can be applied 
to any concentrated flux distribution.  If the flux were concentrated into 
a single clump, for example, it would account for even less than half of 
the total flux.  Nor is the location of the clumpy emission restrictive.  
The interferometric data effectively rule out clumpy emission within the 
30'' primary beam of the SMA at 880\,$\mu$m and within the 
$\sim$30''$\times$45'' approximately elliptical region covered by the 
3-point CARMA mosaic.

It is unlikely that the clumpy structure detected by \citet{koe01} and 
\citet{wil02} could have redistributed itself in the $\sim$10-year interval 
since those observations.  Resonant patterns, like those described by 
W03, change and move on timescales comparable to the orbital 
period of the planet inducing the resonance.  For the putative Neptune-mass 
planet located 65\,AU from the star described in W03, the orbital 
period is approximately 300\,yr.  This is too long to account for a 
large-scale redistribution of material on a 10-year timescale (moreover, 
resonant structures are typically long-lived; they are not expected to 
dissolve on orbital timescales, much less fractions of an orbit).  Stellar 
radiation forces can also alter the spatial distribution of dust; however, 
the grains responsible for the observed emission are likely millimeters in 
size, too large to be expelled by radiation pressure, and the 
Poynting-Robertson drag timescale at the 70\,AU radius of the claimed clumpy 
structure is of order 1\,Gyr. The dust distribution could also change because 
of collisions: the millimeter-sized grain population could be removed by 
collisional grinding, or be rendered invisible by growth into larger bodies 
with smaller millimeter-wavelength opacity.  Both destruction and 
agglomeration occur on timescales at least as long as the collisional 
timescale, which can be estimated as $P_{orb}/\tau$, where $P_{orb}$ is 
the orbital period and $\tau$ is the vertical geometric optical depth of 
the disk \citep[e.g.,][]{lag00}.  For a ring whose annular width is comparable 
to its radius, the vertical optical depth is roughly the observed ratio of 
infrared to bolometric luminosity, sometimes denoted $\tau_{IR}$ 
\citep[e.g.,][]{bac04,chi09}.  Since the Vega debris disk exhibits 
$\tau_{IR}$=2$\times$10$^{-5}$ \citep[e.g.,][]{lag00}, the collision timescale 
must be orders of magnitude larger than the orbital timescale.  Finally, one 
can imagine a scenario wherein the clumps observed $\sim$10 
years ago represent a transient burst of dust released from a catastrophic 
collision between larger parent bodies -- dust whose surface brightness is 
today too small to be detected because of spreading of the ejecta. But even 
at extreme ejecta speeds of 1\,km\,s$^{-1}$, dust could only have spread 
across a few AU, which is an order of magnitude smaller than the linear scale 
corresponding to the beam size of the observations; furthermore, spreading 
by Keplerian differential rotation occurs over timescales at least as long 
as the local orbital period of 300\,yr.  In sum, it is difficult to imagine 
how the clumpy structure could have disappeared in the $\sim$10 years that 
have elapsed between observations.

The interferometric data rule out millimeter-wavelength morphologies that
concentrate the brightness on small scales; this includes a narrow belt or
ring that would concentrate the flux radially.  Ring widths of less than 30\,AU
are ruled out at the 3$\sigma$ level by the interferometric data, and widths
of $>$100\,AU are preferred.  One emerging framework for understanding debris
disk structure asserts that in some systems, large particles may be 
concentrated into a narrow ``birth ring'' in which collisions generate small 
grains carried onto broader orbits by effects of stellar radiation 
\citep{str06}.  This theory has had notable success in explaining the 
differing radial extent of millimeter-wavelength and scattered light data, 
for example in the disk around $\beta$ Pictoris, which exhibits
a ratio $\Delta$R/R of $<$0.5 \citep{wil11}.  However, the Vega debris disk 
is one of several for which spatially resolved millimeter data require a broad 
band ($\Delta R / R \gtrsim 1$) rather than a narrow ring of flux \citep[see 
also HD 107146 and HR 8799;][]{hug11}.  At the coarse spatial resolution and 
limited sensitivity of current observations, not all systems fit neatly into 
a birth ring paradigm.  Millimeter-wavelength maps display a variety of 
morphologies relative to scattered light.  

It should also be noted that the absence of significant clumpy structure on
$\lesssim$100\,AU spatial scales does not rule out the presence of planet-mass
companions.  There are several relevant points to consider.  The W03
model only predicts the underlying planetesimal distribution, and does not 
include a detailed treatment of the effects of collisions on the spatial
distribution of millimeter dust grains.  There is some evidence that 
collisional smoothing may wash out evidence of planetary resonances even for 
the large dust grain populations that dominate the emission at millimeter 
wavelengths \citep{kuc10}, and that far more sensitive observations will be
needed to observe the subtle contrast generated by dynamical interactions with
orbiting planets.  Furthermore, taken as a whole, the observational evidence 
so far points to a complex, multi-component debris disk around Vega.  The data 
presented in this paper rule out a centrally concentrated brightness 
distribution (which would indeed produce a detectable flux concentration on 
$<$100\,AU spatial scales) and instead favor an inner radius for the dust 
disk of 20-100\,AU and a width $>$50\,AU.  However, the near-IR interferometry 
\citep{cia01,def11} and a forthcoming re-analysis of the resolved images and 
SED taking into account all available mid- and far-IR data (K.~Su, private 
communication) provide evidence for multiple dust belts in the system, 
including a warm asteroid belt at a radius of a few AU.  The presence of 
multiple dust belts at different radii is suggestive of a planetary system.  
For example, the HR 8799 system is known to host a system of at least four 
planets, whose orbital radii are bracketed by warm and cold dust belts 
inferred from models of the system SED \citep{che09,su09}.

The series of recent observations calling into question claims of clumpy 
structure in millimeter-wavelength debris disks \citep[e.g.,][this 
work]{pie11,hug11} also has unclear implications for future observations.  
While there is little doubt that debris disk structures like warps, rings, 
and brightness asymmetries are tied to the presence of planets in the disk 
\citep[e.g.][]{wya99,hea00}, it may be necessary to achieve substantially 
greater sensitivity levels than initially thought to observe the density 
concentrations induced by orbital resonances in the large-grain dust 
distribution.  Fortunately, such observations are rapidly becoming possible 
with the advent of the ALMA.  The unprecedented sensitivity of ALMA combined 
with the recent discovery of planets orbiting in debris disks \citep{kal08,
mar08,lag10} will provide an exciting testbed for determining the magnitude 
and wavelength dependence of density perturbations induced by planet-disk 
interactions in a range of different systems.  The next-generation MUSTANG2
instrument on the GBT also promises to be a valuable addition to the suite
of northern-hemisphere instruments.  With 25 times greater sensitivity than
the current MUSTANG detector, the structure of the Vega debris disk will 
observable at high fidelity.  MUSTANG2 will also be able to resolve a number 
of other nearby debris disks, opening up a new long-wavelength regime of 
debris disk imaging. 

\section{Summary}

New observations of the Vega system at a range of millimeter wavelengths
support the observations of \citet{pie11}, indicating that previous detections
of clumpy structure in the Vega debris disk were spurious.  We have analyzed
new interferometric data sets from the SMA and CARMA, as well as a single-dish
observation with the GBT, in the context of several models of disk
structure.  We find that the interferometric data in particular require a disk
with a large (20-100\,AU) inner radius and a broad ($>$50\,AU) radial width.
Clumpy structure on $\lesssim$100\,AU scales accounts for less than half of
the total 860\,$\mu$m flux observed by the JCMT \citep{hol98}. The disk must
therefore be dominated by a smooth and largely axisymmetric millimeter flux 
distribution.

\acknowledgments
We thank Mark Wyatt for providing us with his model of the planetesimal
distribution in the Vega system.  The authors would like to thank the 
MUSTANG instrument team from the University of Pennsylvania, NRAO, Cardiff 
University, NASA-GSFC, and NIST for their efforts on the instrument and 
software that have made this work possible.  The GBT data were obtained under 
the auspices of observing program AGBT08C026.  A.~M.~H. is supported by a 
fellowship from the Miller Institute for Basic Research in Science. S. Dicker 
is supported by NSF AST-1007905.  A.~H. acknowledges support from Millennium
Science Initiative, Chilean Ministry of Economy: Nucleus P10-022-F.  E.~C. 
acknowledges support by NSF grant AST-0909210.

\bibliography{ms}

\end{document}